\RequirePackage{fix-cm}
\documentclass[smallextended]{svjour3}       
\smartqed  
\usepackage{graphicx}
\usepackage{bbm}
\usepackage{mathbbol}
\usepackage{braket}
\usepackage{xcolor}
\usepackage{amssymb,amsmath,amsfonts}
\usepackage{qcircuit}
\usepackage{subcaption}
\newcommand{\up}{\uparrow}
\newcommand{\down}{\downarrow}

\begin{document}

\title{High-fidelity state transfer via quantum walks from delocalized states}



\author{Jo\~ao P. Engster \and Rafael Vieira \and Eduardo I. Duzzioni \and Edgard P. M. Amorim}
\authorrunning{J. P. Engster, R. Vieira, E. I. Duzzioni, and E. P. M. Amorim}

\institute{
J. P. Engster   \at Departamento de F\'isica, Universidade Federal de Santa Catarina, 88040-900, Florian\'oplis, SC, Brazil \\     
R. Vieira       \at Departamento de F\'isica, Universidade Federal de S\~ao Carlos, 13565-905, S\~ao Carlos, SP, Brazil \\
E. I. Duzzioni  \at Departamento de F\'isica, Universidade Federal de Santa Catarina, 88040-900, Florian\'opolis, SC, Brazil \\    
E. P. M. Amorim \at Departamento de F\'isica, Universidade do Estado de Santa Catarina, 89219-710, Joinville, SC, Brazil \\
\email{edgard.amorim@udesc.br}
}

\date{Received: date / Accepted: date}

\maketitle

\begin{abstract}

We study the state transfer through quantum walks placed on a bounded one-dimensional path. We first consider continuous-time quantum walks from a Gaussian state. We find such a state when superposing centered on the starting and antipodal positions preserves a high fidelity for a long time and when sent on large circular graphs. Furthermore, it spreads with a null group velocity. We also explore discrete-time quantum walks to evaluate the qubit fidelity throughout the walk. In this case, the initial state is a product of states between a qubit and a Gaussian superposition of position states. Then, we add two $\sigma_x$ gates to confine this delocalized qubit. We also find that this bounded system dynamically enables periodic recovery of the initial separable state. We outline some applications of our results in dynamic graphs and propose quantum circuits to implement them based on the available literature.

\end{abstract}


\section{Introduction} \label{sec.1}

The main problem of communication is to reproduce at one point, the same or approximately the information from another point. This statement pointed out by C. Shannon \cite{shannon1948mathematical} is also a fundamental issue in the quantum context. Any promising platform for quantum information processing should faithfully store, transfer, and recover the information from quantum states \cite{nielsen2010quantum}. Quantum walks in their discrete- \cite{aharonov1993quantum} and continuous-time \cite{farhi1998quantum} versions have brought a novel theoretical perspective for performing quantum tasks \cite{kempe2003quantum,venegas2012quantum}. For instance, they have been studied as a framework for building quantum search engine algorithms \cite{portugal2013quantum} and a possible route to realizing universal quantum computation \cite{childs2009universal,lovett2010universal}. Moreover, there are many current experimental setups to implement them \cite{wang2013physical,flamini2019photonic}.

Continuous-time quantum walks (CTQWs) are based on continuous-time Hamiltonian, and discrete-time quantum walks (DTQWs) have their dynamical evolution dictated by a unitary transformation composed of a quantum coin and a conditional displacement operator at discrete time steps. In CTQWs, as a given state evolves, it spreads to all the vertices according to the graph geometry and the hopping rates between vertices. Eventually, it reemerges in another place with or without loss of information. This process is called state transfer. Although CTQWs have perfect state transfer in a few small graphs, it is limited for large graphs. There are many relevant studies about this subject \cite{aharonov2001quantum,ahmadi2003mixing,kendon2011perfect,godsil2012state,barr2014periodicity,dutta2022perfect}, mainly dealing with a walk whose initial state is localized on just one vertex (position). In such a case, the state transfer is only the probability of reaching a specific target vertex. However, we can change the localized initial condition by delocalizing the initial state over positions. In this scenario, the fidelity between the states around the initial and target vertices becomes an appropriate figure of merit to evaluate the state transfer since it includes how the state spreads to the vertices \cite{vieira2021quantum}.

The delocalization of the initial state weighted according to a distribution function was employed in some experimental platforms using a photon as walker \cite{cardano2015quantum,su2019experimental,derrico2020two}, theoretical works on quantum walks \cite{vieira2021quantum,abal2006quantum,abal2006quantumE,valcarcel2010tailoring,romanelli2010distribution,vieira2013dynamically,vieira2014entangling,zhang2016creating,orthey2017asymptotic,orthey2019connecting,ghizoni2019trojan,orthey2019weak,khalique2021controlled}, and spin chains protocols \cite{Nikolopoulos2004,Osborne2004,Plenio2004,Haselgrove2005,Plenio2005,Shi2005,Karbach2005,Chen2006,Hartmann2006,Bose2007,Banchi2010,Nicacio2016,Moradi2019,dutta2023quantum}. Starting a DTQW from a delocalized qubit leads to new phenomena that affect the transport and entanglement between coin and position states of a walker. For instance, a quantum walker from a local state has the same limit velocity regardless of the quantum coin used. However, when it begins from a Gaussian state, its limit velocity depends strongly on the coin phases and the initial dispersion of the state \cite{orthey2019connecting}. Only two qubits that start a quantum walk from a single position exhibit maximal entanglement asymptotically, but a continuous set of delocalized qubits lead to maximal entanglement \cite{orthey2017asymptotic}. When one $\sigma_x$ gate (Pauli-X) is used as the quantum coin among Hadamard ones on a lattice, an initial broad Gaussian state evolves into a trojan wave packet \cite{ghizoni2019trojan}. Furthermore, the joint of these two gates at specific times and positions throughout the walk yields the corralling of this quantum state, preventing its spread even better \cite{vieira2021quantum}. 

When a quantum walk occurs on a cycle graph, the initial state spreads with null group velocity \cite{kempf2009group} and superposes at the periodic boundary. Due to this superposition, it may return to its original shape centered at the antipodal vertex, which can be far away from its initial position. Therefore, by taking a delocalized initial state, we aim to evaluate this dynamical evolution as one potential means of achieving high-fidelity state transfer using CTQWs on cycle graphs. Moreover, we examine DTQWs that start from a delocalized qubit to assess whether the qubit, acting as an information carrier, is also preserved along the walk. 

The article is structured as follows. In Sect.~\ref{sec:2}, we briefly introduce the mathematical formalism of CTQWs. In Sect.~\ref{sec:3}, we review the CTQWs starting from a local state, obtain an expression for the fidelity of delocalized states over time, and study such walks starting from Gaussian states confined for a long time and over large graphs. In Sect.~\ref{sec:4}, we review the DTQWs by comparing their dynamical evolution starting from one vertex as long as the state becomes a broad Gaussian one. In Sect.~\ref{sec:5}, we discuss an application of our results in the context of dynamic graphs and quantum circuits. Finally, Sect.~\ref{sec:6} presents some concluding remarks.

\section{Continuous-time quantum walks}\label{sec:2}

CTQWs are usually expressed as Markov processes, whose time evolution is driven by the Schr\"{o}dinger equation
\begin{equation}
i\hbar\frac{d}{dt}\ket{\Psi(t)} = H\ket{\Psi(t)}.
\label{schrodinger}
\end{equation}
From now, we assume $\hbar=1$, and by inserting the completeness relation for $\ket{b}$, we have
\begin{equation}
i\frac{d}{dt}\braket{a|\Psi(t)}=\sum_b\braket{a|H|b}\braket{b|\Psi(t)},
\label{Schrodinger}
\end{equation}
in which $H$ is characterized with respect to a graph $\mathcal{G}$. The graph $\mathcal{G}=(\mathcal{V},\mathcal{E})$ is given by a finite and discrete Hilbert space $\mathcal{V}$ spanned by $\{\ket{a}\}$ with $a\in\mathbb{Z}$ corresponding to $n$ vertices, while the set $\mathcal{E}$ of edges specifies which pairs of vertices are connected. The graph geometry constrains the particle, allowing its movement only between connected vertices. The Hamiltonian $H$ is  
\begin{equation}
H_{ab}=
\begin{cases}
\gamma d_a   &\text{for}\quad a=b\\
-\gamma      &\text{for}\quad a\neq b\, |\,\left(a,b\right)\in \mathcal{E}\\
0            &\text{for}\quad a\neq b\, |\,\left(a,b\right)\not\in \mathcal{E},
\end{cases}
\label{Hamiltonian}
\end{equation}
where $d_a$ (degree) is the number of edges connected to the vertex $a$, and $\gamma$ is the hopping rate per time from the vertex $a$ to $b$ and vice versa. Let us consider $\gamma=1$ throughout this work. Therefore, CTQWs evolve over time by 
\begin{equation}
\ket{\Psi(t)}=U\ket{\Psi(0)},
\label{Psi_t}
\end{equation}
with $U=e^{-iHt}$ being the unitary time evolution operator. After diagonalizing $U$, we obtain
\begin{equation}
\ket{\Psi(t)}=\sum_{b=0}^{n-1} e^{-i\lambda_bt}\braket{\Phi_b|\Psi(0)}\ket{\Phi_b},
\label{Psi_t2}
\end{equation}
where $\lambda_b$ are the corresponding eigenvalues of the orthonormal eigenvectors $\ket{\Phi_b}$ and the sum above scales according to the number of vertices $n$ of the graph. 

\section{CTQW on the cycle graph}\label{sec:3}

Let us consider a full-cycle $C_n$ on $n$ vertices whose Hamiltonian is $H=A-2\mathbb{1}$, where $A$ is a circulant matrix and $\mathbb{1}$ is the identity matrix (order $n$). Since both matrices commute, by neglecting an irrelevant phase, the time evolution operator gives us $U=e^{-iAt}$. The circulant matrix $A$ can be diagonalized by the Fourier matrix, 
\begin{equation}
F=\frac{1}{\sqrt{n}}  
\left(
\begin{matrix}
1      & 1            & 1               & \cdots & 1               \\
1      & \omega       & \omega^2        & \cdots & \omega^{n-1}    \\ 
1      & \omega^2     & \omega^4        & \cdots & \omega^{2(n-1)} \\
\vdots & \vdots       & \vdots          & \ddots & \vdots          \\
1      & \omega^{n-1} & \omega^{2(n-1)} & \cdots & \omega^{(n-1)^2}
\end{matrix}\right),
\label{Fmatrix}
\end{equation}
with $\omega = e^{2\pi i/n}$. Then, the matrix $FAF^{\dagger}$ is diagonal. Let us denote the $j$-th column vector of $F$ concerning the graph vertices as 
\begin{equation}
\ket{F_j}=\frac{1}{\sqrt{n}}\ket{\omega_j}=\frac{1}{\sqrt{n}} \sum_{a=0}^{n-1} \omega^{ja}\ket{a},
\label{Fj}
\end{equation}
which composes an orthonormal basis, such that 
\begin{equation}
\mathbb{1}=\sum_{j=0}^{n-1} \ket{F_j}\bra{F_j}=\frac{1}{n} \sum_{j=0}^{n-1} \ket{\omega_j}\bra{\omega_j}.
\label{Identity}
\end{equation}
Therefore, $\ket{F_j}$ are the eigenvectors and
\begin{equation}
\lambda_j=\omega^j+\omega^{j(n-1)}=2\cos\left(\frac{2\pi j}{n}\right) 
\label{Eigenvalues_CTQW}
\end{equation}
are the corresponding eigenvalues of $A$ \cite{ahmadi2003mixing}. In quantum computing language, the matrix $F$ implements the quantum Fourier transform (QFT). It acts on a basis state $\ket{j}$ similarly to Eq. (\ref{Fj}) as
\begin{equation}
QFT\ket{j}=\frac{1}{\sqrt{n}} \sum_{a=0}^{n-1} \omega^{ja}\ket{a}.
\label{QFT}
\end{equation}
The action of the inverse of the QFT, $IQFT$, is determined by
\begin{equation}
IQFT\ket{a}=\frac{1}{\sqrt{n}} \sum_{j=0}^{n-1} \omega^{-ja}\ket{j}.
\label{IQFT}
\end{equation}
Eqs. (\ref{QFT}) and (\ref{IQFT}) will be used later to build quantum circuits that apply to simulations of quantum walks in quantum computers.

\subsection{Initial local state}

\begin{figure}[h!]
\centering
\includegraphics[width=0.6\linewidth]{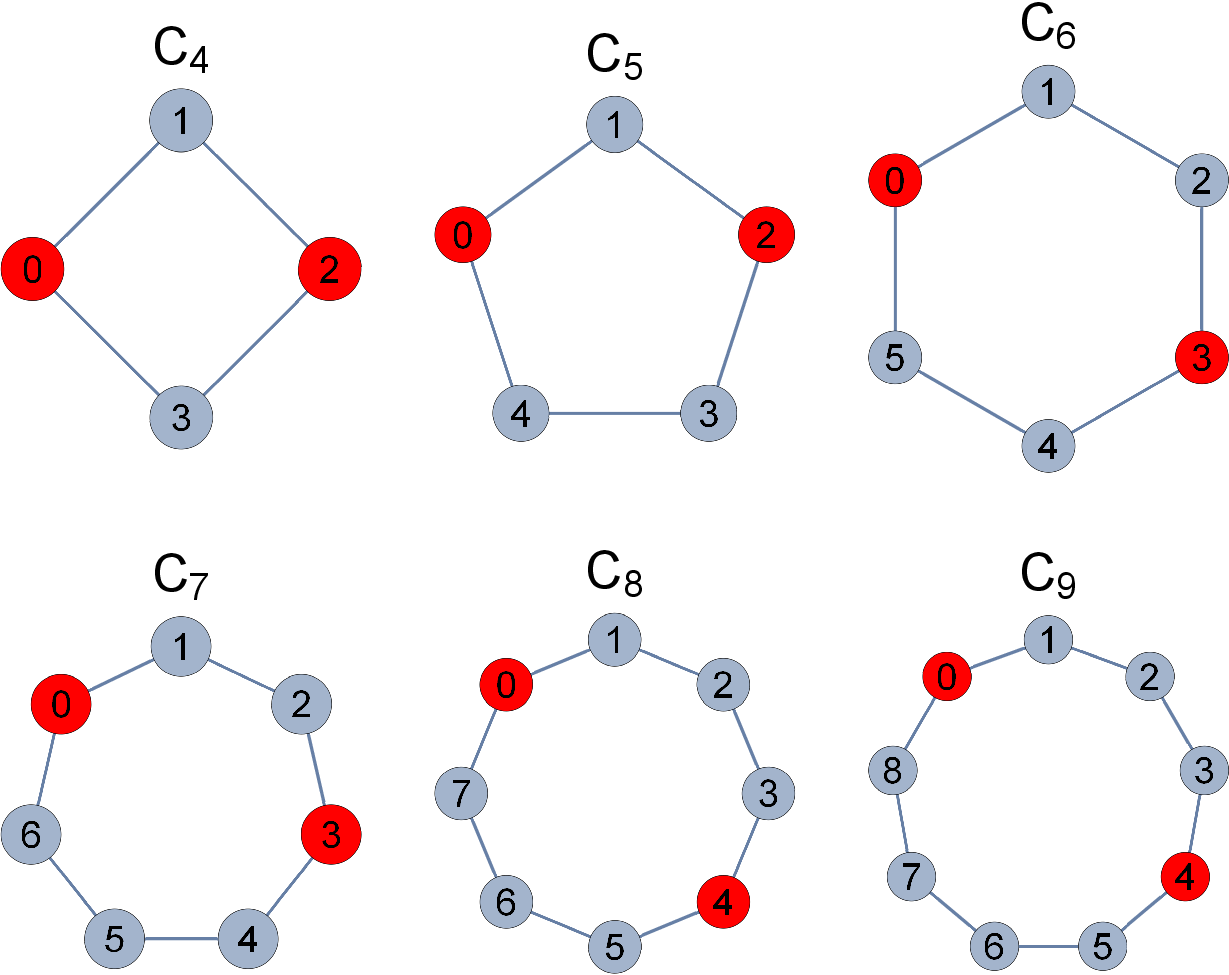}
\caption{Graphs $C_n$ from $n=4$ to $n=9$. The initial vertex $a=0$ and antipodal vertex $b=[n-(n~\text{mod}~2)]/2$ in red.}
\label{fig.1}
\end{figure}

Let us take a localized state $\ket{\Psi(0)}=\ket{0}$ on $C_n$ such that $\ket{\Psi(t)} = e^{-iAt}\ket{0}$, and by using Eqs.~(\ref{Fj})–(\ref{Eigenvalues_CTQW}) we get
\begin{equation}
\ket{\Psi(t)} = \frac{1}{n} \sum_{j=0}^{n-1} e^{-2i\cos \left(\frac{2\pi j}{n}\right)t} \ket{\omega_j}.
\end{equation}
Since our target vertex is the farthest one from $a$, i.e., the antipodal vertex $b$ (see Fig. \ref{fig.1}), then the probability $\mathcal{P}_b(t)= \left|\braket{b|\Psi(t)}\right|^2$ can be written as
\begin{equation}
\mathcal{P}_b(t)=\frac{1}{n^2} \left|\sum_{j=0}^{n-1} e^{-2i\left[\cos \left(\frac{2\pi j}{n}\right)t-\frac{\pi j}{n}b\right]}\right|^2.
\label{Prob_b}
\end{equation}

\begin{figure}[h!]
\centering
\includegraphics[width=0.6\linewidth]{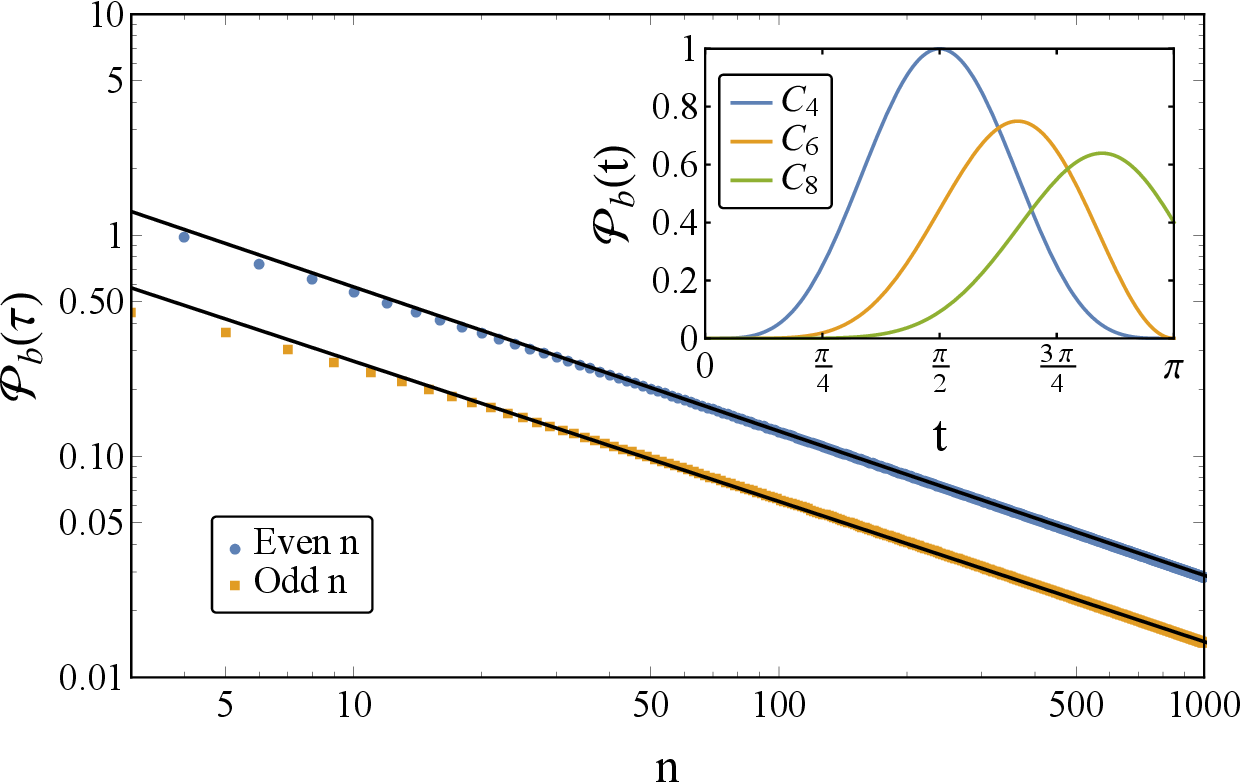}
\caption{Probability at the antipodal vertex $b$ for the transfer time $\tau$. $\mathcal{P}_b(\tau)\propto n^{-\alpha}$ have exponents $\alpha=0.6524\pm0.0004$ and $0.6344\pm0.0009$, respectively, for even and odd values of $n$, while $\tau\sim n/4$. Inset: Probabilities at $b$ over time for the cycle graphs $C_4$, $C_6$ and $C_8$ (see Eqs.~(\ref{Prob_b_C4_6_8})) showing the first maximum of these functions at $t=\tau$.}
\label{fig.2}
\end{figure}

It is worth noticing that since we are dealing with a local state, the probability in Eq.~(\ref{Prob_b}) also corresponds to the fidelity over time between the initial state and the one at the antipodal vertex. For instance, for the cycle graphs $C_4$, $C_6$ and $C_8$ such that $b=2$, $3$, and $4$, respectively, we have 
\begin{align}
\mathcal{P}_2(t) &= \sin^4t, \nonumber \\
\mathcal{P}_3(t) &= \frac{16}{9} \sin ^4\left(\frac{t}{2}\right) \sin^2t, \nonumber \\
\mathcal{P}_4(t) &= \frac{1}{16} \left[1+\cos(2t)-2 \cos \left(\sqrt{2}t\right)\right]^2,
\label{Prob_b_C4_6_8}
\end{align}
showing that $C_4$ exhibits perfect local state transfer for $t=(2l+1)\frac{\pi}{2}$ with $l\in\mathbbm{N}$ where $t=\pi/2$ is the transfer time $\tau$ to reach the first local maximum of $\mathcal{P}_b(t)$ as shown in the inset of Fig.~\ref{fig.2}. Figure~\ref{fig.2} shows how the probabilities at the antipodal vertex $b$ follow a power law such as $\mathcal{P}_b(\tau)\propto n^{-\alpha}$. This result inevitably implies that the local state transfer gets worse as the number of vertices grows.

\subsection{Initial delocalized states} \label{SubSec:InitDelocState}

Let us consider an initial state delocalized over many vertices that follow a certain distribution function $f(j)$. This delocalized state can be written as a superposition from a subset of vertex states of $\mathcal{G}$ as
\begin{equation}
\ket{\Psi(0)}=\sum_{j=-a}^{a} f(j)\ket{j+a},
\label{Delocalizedstate}
\end{equation}
where $f(j)$ is centered at the origin and satisfies the normalization condition $\sum_j|f(j)|^2=1$. This condition is assured because the number of vertices in the graph, denoted by $n=2a+1$ is much larger than the initial standard deviation $s$ of the quantum state. Now, by using the same steps in the previous section, we arrive at
\begin{equation}
\mathcal{P}_b(t)=\frac{1}{n^2}\left|\sum_{j,l=0}^{n-1}\frac{f(j-a)}{e^{2i\left[\cos \left(\frac{2\pi l}{n}\right)t-\frac{\pi l}{n}(b-j)\right]}}\right|^2,
\label{Prob_b_Gaussian}
\end{equation}
where the sum in $j$ was extended for the whole graph. Taking all values of $\mathcal{P}_b(t)$ from $b=0$ to $n-1$ gives us the probability distribution over the graph vertices at a specific time $t$. 

\subsection{Fidelity of delocalized states}

Fidelity is commonly used as a metric to quantify the similarity between two quantum states \cite{vieira2021quantum}. Particularly, the fidelity between the initial state $\ket{\Psi(0)}$ centered at vertex $j=a$ and the time-evolved state $\ket{\Psi(t)}$ at time $t$ and centered at $j=b$ evaluates the state transfer quality whenever $a\neq b$, and the periodicity for $a=b$. Since these states could be on different vertices, we use an operator $\mathcal{D}=\sum_j \ket{j+b-a}\bra{j}$ that acting on the initial state $\mathcal{D}\ket{\Psi(0)}$ displaces the center of $\ket{\Psi(0)}$ from $j=a$ to $b$. Then, the fidelity $\mathcal{F}(t)=|\braket{\Psi(0)|\mathcal{D}^\dagger|\Psi(t)}|^2$ becomes
\begin{equation}
\mathcal{F}(t)=\frac{1}{n^2}\left|\sum_{j,j',l=0}^{n-1} \frac{f(j-a)f(j'-a)}{e^{2i\left[\cos \left(\frac{2\pi l}{n}\right)t-\frac{\pi l}{n}(j-j'+b)\right]}}\right|^2,
\label{Fidelity}
\end{equation}
ranging from $\mathcal{F}=0$ for two orthogonal states until $\mathcal{F}=1$, for two identical states up to an overall phase factor. Observe that if the distribution $f(j)$ were a delta function $\delta(j)$ we would recover Eq.~(\ref{Prob_b}) as expected. 

\subsection{Gaussian states}

\begin{figure}[h!]
\centering
\includegraphics[width=0.6\linewidth]{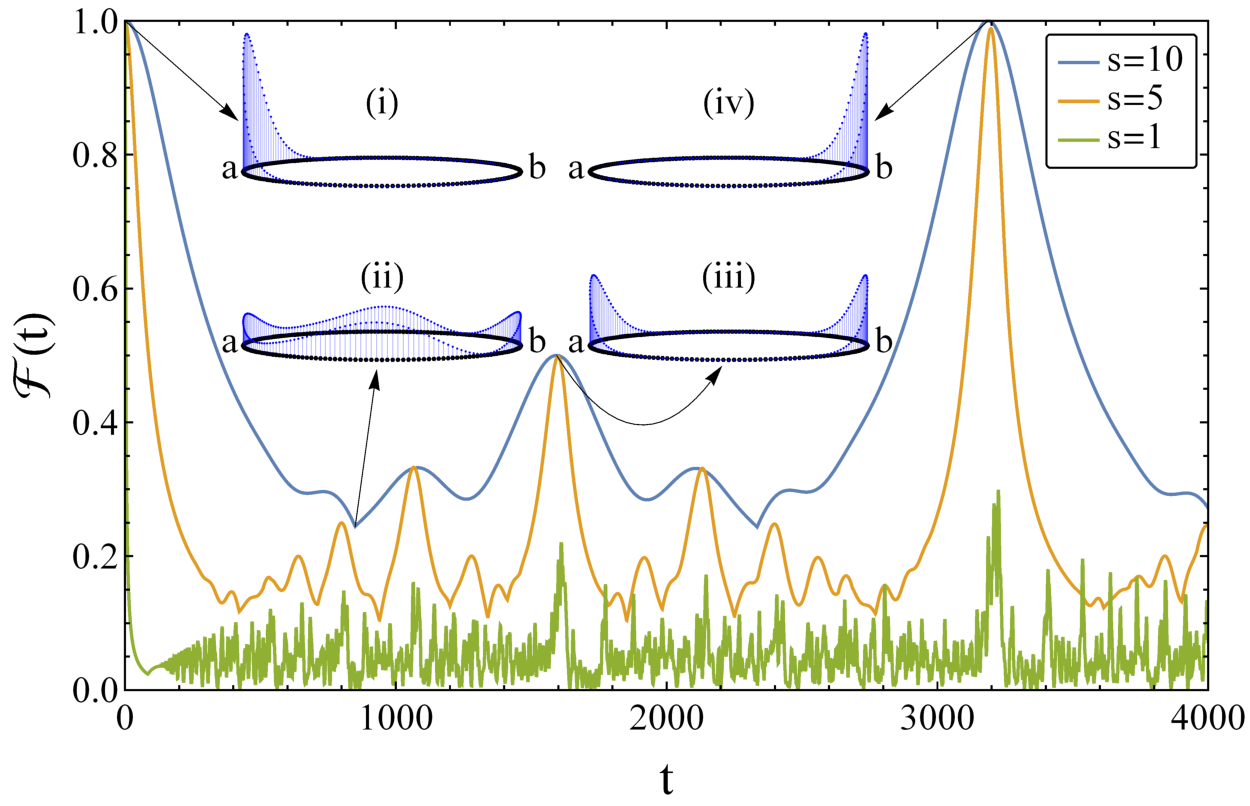}
\caption{Fidelity over time for CTQW on $C_{200}$ starting from Gaussian states with initial dispersion $s=1$ (bottom), $5$ and $10$ (top). The insets show the time evolution of the probability distribution of the Gaussian state ($s=10$). It starts as (i) a Gaussian state centered at $a$, then (ii) spreads to all the vertices, next evolves to (iii) a superposition of two balanced Gaussian distributions centered at $a$ and $b$, and finally returns to (iv) the original shape centered at the antipodal vertex $b$ after a transfer time $\tau$.}
\label{fig.3}
\end{figure}

Let us consider a state whose distribution function is a Gaussian one centered at vertex $a$
\begin{equation}
\ket{\Psi(0)}=\sum_{j=0}^{n-1} f(j-a)\ket{j}=\sum_{j=0}^{n-1} \frac{e^{-\left(\frac{j-a}{2s}\right)^2}}{(2\pi s^2)^{\frac{1}{4}}}\ket{j},
\label{Gaussianstate}
\end{equation}
where $s$ is the initial standard deviation (dispersion). Figure~\ref{fig.3} shows the fidelity $\mathcal{F}(t)$ for CTQWs over a $C_{200}$ graph starting the walks from Gaussian states centered at $a$ with a few initial values of $s$. The states spread over all the vertices of the graph, then after a transfer time $\tau$, the superposition pattern of the wave packet returns to the initial Gaussian state, but now it is centered at the antipodal vertex $b$. Regarding this process as a possible way to get a state transfer, it is notable the dependence on the initial dispersion of the state. While we get $\mathcal{F}(\tau)\approx1$ for large $s$, indicating an almost perfect transfer, this value drops to around $0.30$ for $s=1$. The insets of Fig.~\ref{fig.3} present the dynamical evolution of the transfer of the wider Gaussian state through some frames of the probability distribution at certain times. They start from the inset (i), which shows the initial state $\ket{\Psi(0)}$ centered at vertex $a$ at $t=0$. The inset (ii) detaches one of the local minima of the fidelity when the state is closer to a uniform distribution over the vertices. The inset (iii) corresponds to half of the transfer time when the state achieves the maximum spreading as a superposition 
\begin{equation}
\ket{\Psi(\tau/2)}=\sum_{j=0}^{n-1} \frac{f(j-a)+f(j-b)}{\sqrt{2}} \ket{j},
\label{Psi_tau/2}
\end{equation}
with $\mathcal{F}(\tau/2)\approx0.50$. The last inset (iv) shows the time-evolved state centered at antipodal vertex $b$ after being transferred at $t=\tau$.

\begin{figure}[h!]
\centering
\includegraphics[width=0.6\linewidth]{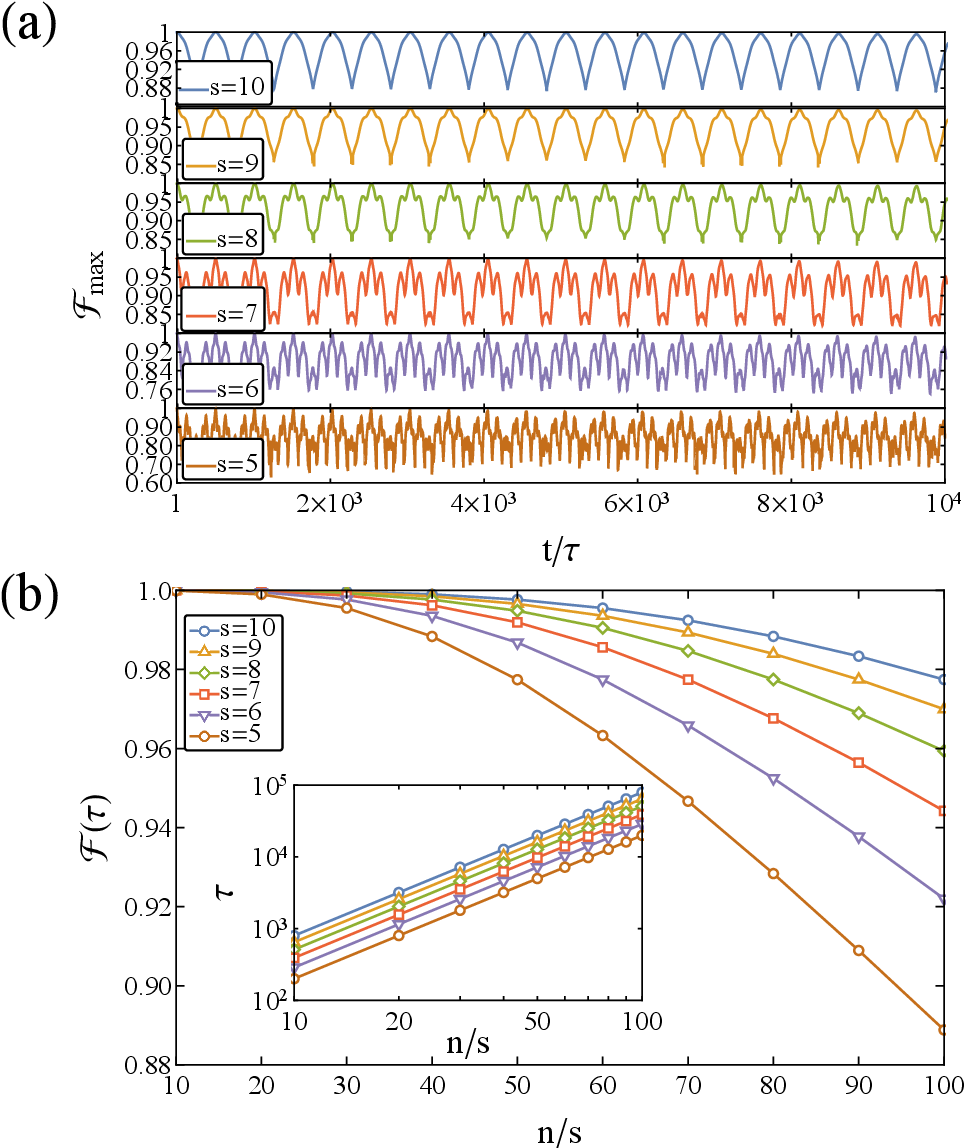}
\caption{Transfer of Gaussian states for long times and on large graphs via CTQWs: (a) Local maxima of fidelity from $s=5$ (bottom) to $10$ (top) over $C_{100}$. After establishing the transfer time $\tau$, we seek the value of the next fidelity peak within the range $[t-\tau/2, t+\tau/2]$ from $t=2\tau$ up to $t=10^4\tau$ to proper identify when the states have their center at the vertex $a$ or $b$. (b) Fidelity at $t=\tau$ over the number of standard deviations $n/s$ from  $s=5$ (bottom) to $10$ (top). Inset: transfer time grows as the power law $\tau\sim 0.08 n^2$. The solid lines between points are just guides for the eyes.}
\label{fig.4}
\end{figure}

CTQWs starting from a Gaussian state reach at least one almost perfect state transfer whose quality is subject to the initial standard deviation of the state. To address how robust is this state transfer for a longer time, Fig.~\ref{fig.4} (a) shows all the local maxima of fidelity for CTQWs starting from Gaussian states on $C_{100}$ from the first transfer time $\tau$ up to $10^4\tau$. The state transfer of a local state has a small and non-periodic fidelity over large cycle graphs. At this point, it becomes clear that a Gaussian state transfer outperforms a local state transfer once the local maxima of fidelity exhibit a quasi-periodic behavior with a high-valued upper-bound limit. This upper-bound fidelity drops slowly and remains higher than $0.99$ up to $t\sim10^4\tau$ for large $s$. So, this indicates a remarkable preservation of the quantum state fidelity, whereas, for the local state, the same upper-bound fidelity is lower than $0.5$. 

To investigate how the fidelity of the first state transfer behaves over great distances, Fig.~\ref{fig.4} (b) shows the fidelity of Gaussian states at $t=\tau$ on graphs with $n$ varying from $10$ up to $100$ times their corresponding initial standard deviations. The broader initial state has the best state transfer, whatever the distance between the initial and antipodal vertices. It shows that the state transfer depends on the initial delocalization of the state, regardless of the graph size. Notice that the transfer time $\tau$ is quadratically greater for all Gaussian states than the local state, with all states having ballistic spreading. It means that the state dispersion over time follows $\sigma\propto t$. Recalling the uncertainty principle, the smaller the initial state dispersion in position, the larger the wave packet spreads. Thus, the localized state spreads the fastest, with a rate of $d\sigma/dt=\sqrt{2}$, while Gaussian states spread more slowly as their initial dispersion increases, resulting in lower transfer rates. Furthermore, besides the delocalization, the smoothness of the state is also a significant feature in achieving high fidelity (see Appendix \ref{sec:apend}). 

\section{Discrete-time quantum walks}\label{sec:4}

The DTQW model here involves a dynamical evolution with one additional degree of freedom (qubit) compared to the CTQW model. The DTQW describes a walk of a spin-$1/2$ particle over a one-dimensional lattice whose dynamical evolution is driven by a unitary operator at discrete-time steps. Formally, a DTQW state belongs to a Hilbert space $\mathcal{H}=\mathcal{H}_C\otimes\mathcal{H}_P$, where $\mathcal{H}_C$ is the coin space, a complex two-dimensional vector space spanned by $\left\{\ket{\up}, \ket{\down}\right\}$, and $\mathcal{H}_P$ is the position space, a numerable infinite-dimensional vector space spanned by $\left\{\ket{j}\right\}$ whose integer $j$ is a discrete position (vertex) on a regular lattice. Thus an initial quantum walk state is given by
\begin{eqnarray}
\ket{\Psi(0)}&=& \sum_j\left[a^{\up}_j(0)\ket{\up}+a^{\down}_j(0)\ket{\down}\right]\otimes \ket{j} \nonumber \\
&=&\left[\cos\alpha\ket{\up}+e^{i\beta}\sin\alpha\ket{\down}\right]\otimes\sum_j f(j)\ket{j}, 
\label{Psi_0}
\end{eqnarray}
where $a^{\up}_j(t)$ and $a^{\down}_j(t)$ for $t=0$ are the initial amplitudes of spin up and down, respectively, with the sum in $j$ being overall positions on the lattice, $\alpha\in[0,\pi/2]$ is the half polar angle, and $\beta\in[0,2\pi]$ is the azimuthal angle in the Bloch sphere representation \cite{nielsen2010quantum}. The function $f(j)$ is the initial distribution function and $\sum_j|f(j)|^2=1$ is the normalization condition. We use here $f(j)$ as a delta function (local state) and a discrete Gaussian distribution with $s$ being the initial position dispersion of the state, following the same way described in the CTQWs section above.

The state after $t_N$ discrete-time steps is given by
\begin{equation}
\ket{\Psi(t_N)}=\mathcal{T}\prod_{t=1}^{t_N} U(j)\ket{\Psi(0)},
\label{Psi_n}
\end{equation}
where $\mathcal{T}$ specifies a time-ordered product and $U(j)=S[C(j)\otimes\mathbbm{1}_P]$ is the time evolution operator composed by the identity operator $\mathbbm{1}_P$ in $\mathcal{H}_P$, a position-dependent quantum coin $C(j)$ belonging to $SU(2)$, and the conditional displacement operator $S$. We employ the Hadamard gate
\begin{equation}
H=\frac{1}{\sqrt2}[\ket{\up}\bra{\up}+\ket{\up}\bra{\down}+\ket{\down}\bra{\up}-\ket{\down}\bra{\down}],
\label{Hadamard}
\end{equation}
and NOT gate (Pauli-X),
\begin{equation}
\sigma_x=\ket{\up}\bra{\down}+\ket{\down}\bra{\up},
\label{Sigmax}
\end{equation}
as quantum coins. The quantum coin acts on the qubit by setting it on a new superposition of spin states, for example, $H\ket{\up}=(\ket{\up}+\ket{\down})/\sqrt{2}$ or $\sigma_x(\ket{\up}+e^{i\theta}\ket{\down})/\sqrt{2}=(e^{i\theta}\ket{\up}+\ket{\down})/\sqrt{2}$. At last, the conditional displacement operator 
\begin{equation}
S=\sum_j(\ket{\up}\bra{\up}\otimes\ket{j+1}\bra{j}+\ket{\down}\bra{\down}\otimes\ket{j-1}\bra{j})
\label{DisplaceOp}
\end{equation}
displaces each spin component in opposite directions, i.e., it shifts the amplitude of spin up (down) to the right (left) neighbor position, consequently entangling spin and position. The time-evolved state $\ket{\Psi(t)}$ is calculated by an iterative procedure whose recurrence equations can be obtained from Eq.~(\ref{Psi_n}) through $\ket{\Psi(t)}=U(j)\ket{\Psi(t-1)}$ \cite{vieira2014entangling} with Eqs.~(\ref{Hadamard})–(\ref{DisplaceOp}). The total probability over a position $j$ is given by the sum between the spin up and down components,
\begin{equation}
\mathcal{P}_j(t)=|\left(\bra{\up}\otimes\bra{j}\right)\ket{\Psi(t)}|^2+|\left(\bra{\down}\otimes\bra{j}\right)\ket{\Psi(t)}|^2=|a^{\up}_j(t)|^2+|a^{\down}_j(t)|^2,
\label{P_jt}
\end{equation}
and the standard deviation over time is
\begin{equation}
s(t)=\sqrt{\sum_j \left\{j^2\mathcal{P}_j(t)-\left[j\mathcal{P}_j(t)\right]^2\right\}}.
\label{st}
\end{equation}
Since $\ket{\Psi(t)}$ is pure over time, the entanglement is carried out by the von Neumann entropy,
\begin{equation}
S_E(\rho(t))=-\mathrm{Tr}[\rho_C(t)\log_2\rho_C(t)]
\label{Entropy}
\end{equation}
where $\rho_C(t)=\mathrm{Tr}_P[\ket{\Psi(t)}\bra{\Psi(t)}]$ is the partially reduced coin state \cite{bennett1996concentrating} and $\mathrm{Tr}_P[\cdot]$ is the trace over the positions. Therefore
\begin{equation}
\rho_C(t)=A^{\up}(t)\ket{\up}\bra{\up}+M(t)\ket{\up}\bra{\down}+M^*(t)\ket{\down}\bra{\up}+A^{\down}(t)\ket{\down}\bra{\down}
\label{RhoC}
\end{equation}
where $A^{\up}(t)=\sum_j|a_j^{\up}(t)|^2$, $M(t)=\sum_j a_j^{\up}(t)a_j^{\down *}(t)$ with $z^*$ being the complex conjugate of $z$, and $A^{\down}(t)=\sum_j|a_j^{\down}(t)|^2=1-A^{\up}(t)$. After diagonalizing $\rho_C(t)$, we obtain the following eigenvalues
\begin{equation}
\Lambda_{\pm}(t)=1/2\pm\sqrt{1/4-A^{\up}(t)(1-A^{\up}(t))+|M(t)|^2},
\label{Eigenvalues}
\end{equation}
which allows us to write
\begin{equation}
S_E(t)=-\Lambda_+(t)\log_2\Lambda_{+}(t)-\Lambda_{-}(t)\log_2\Lambda_{-}(t)
\label{SE(t)}
\end{equation}
such that $S_E$ is null for separable states up to $1$ for maximal entanglement between spin and position.

\subsection{Bounded DTQW}

The asymptotic properties regarding the dynamics and entanglement of a DTQW can be analytically derived using Fourier analysis, since the time-evolution operator is diagonal in the dual $k$-space \cite{vieira2021quantum,abal2006quantum,orthey2017asymptotic,orthey2019connecting,ambainis2001one}. Particularly, numerical and analytical calculations have shown that a Hadamard walk whose initial state is a Gaussian one with a large dispersion splits into two Gaussian wave packets moving in opposite directions \cite{vieira2021quantum,orthey2019connecting,ghizoni2019trojan}, resulting in
\begin{eqnarray} 
\ket{\Psi(t)}&=&\ket{\psi_+}\otimes \sum_jf(j-t/\sqrt{2})\ket{j} \nonumber \\
&& + (-1)^t|\ket{\psi_-}\otimes\sum_jf(j+t/\sqrt{2})\ket{j},
\label{Psit}
\end{eqnarray}
where 
\begin{eqnarray}
\ket{\psi_{\pm}}&=&\frac{\left(1\pm \sqrt{2} \right)\cos \alpha+e^{i \beta}\sin \alpha}
{2\left(2\pm \sqrt{2}\right)}
\left(
\begin{matrix}
1\pm\sqrt{2}\\
1
\end{matrix}\right),
\end{eqnarray}
are orthogonal states and $f(j\pm t/\sqrt{2})$ are Gaussian distributions centered at $j=\mp t/\sqrt{2}$ throughout the walk. Notice that this result assumes a small time scale \cite{vieira2021quantum}, and the relative velocity between split wave packets depends on the quantum coin used \cite{orthey2019connecting}.

Here, we explore DTQWs driven by Hadamard coins constrained by two $\sigma_x$ gates as quantum coins placed at opposite positions. The walks start from a separable state such as Eq.~(\ref{Psi_0}) whose qubit is positioned on $j=0$ (local state) or spread following a discrete Gaussian distribution. Each numerical realization of the DTQW begins from a qubit $(\alpha,\beta)=(0,0)$ up to $(\pi/2, 2\pi)$ with independent increments of $0.05$ for $\alpha$ and $0.1$ for $\beta$. Then, we carry out an average over a set of $2,016$ qubits for each time step during their walks, as in earlier studies. \cite{vieira2013dynamically,vieira2014entangling,orthey2017asymptotic,orthey2019connecting,ghizoni2019trojan}. 

\begin{figure}[h!]
\centering
\includegraphics[width=\linewidth]{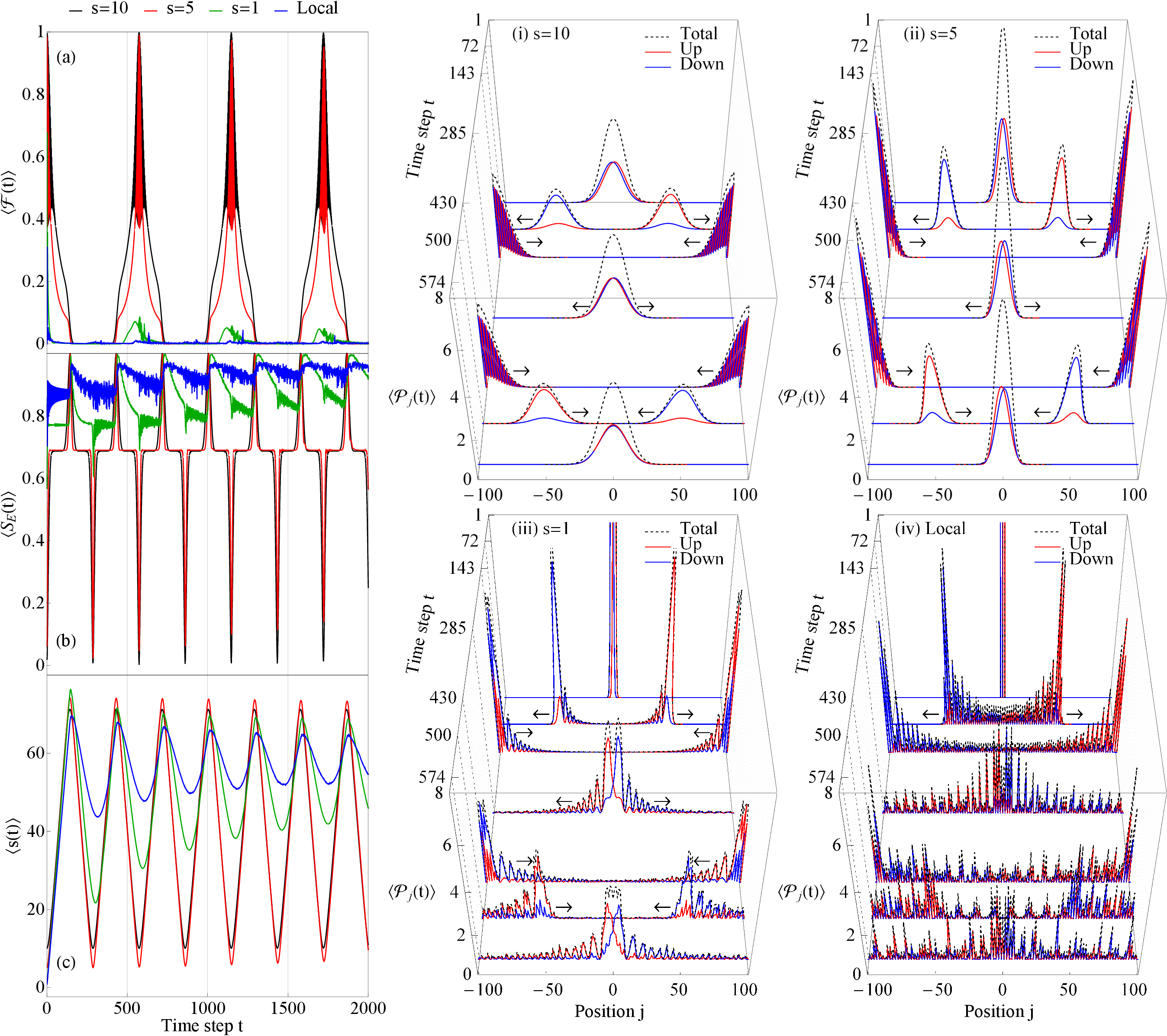}
\caption{Confining a qubit via DTQWs: Average values of (a) fidelity, (b) entanglement, and (c) standard deviation over time of quantum walks starting from Gaussian states centered at $j=0$ with initial dispersion $s=10$ (black), $5$ (red), $1$ (olive), and a localized state (blue). The panels (i)-(iv) show the total probability distribution over time (dashed black line) and its up (red) and down (blue) spin components for all cases. The quantum walk states are confined within $j=\pm 100$ by two $\sigma_x$ gates represented by the gray opposite walls. The arrows indicate the direction of displacement of the wave packets.} 
\label{fig.5}
\end{figure}

Figure~\ref{fig.5} shows the average probability over time of such walks starting from a (i) localized state on $j=0$ and (ii)–(iv) Gaussian states centered at $j=0$. Note that, Gaussian states are defined within $j=\pm L$, such that the ratio $s/L$ are $1\%$, $5\%$, and $10\%$, respectively. The insets (a)–(c) show that both local and Gaussian ($s=1$) states spread continually over time, while the other broader Gaussian states recover periodically their respective initial dispersion and two distinct product states ($S_E\approx 0$), such that the second one has high fidelity $\mathcal{F}\approx 1$. It means that, as soon as the walk begins, the spin and position entangle over time. Then, the initial separable state becomes entangled. The wave packet splits into two peaks traveling in opposite directions. The effect of the $\sigma_x$ gates gives rise to a chiral reflection of the wave packet due to the conditional displacement. So, the wave packet reaches the $\sigma_x$ gates, and their spin components swap, i.e., the amplitude of spin up becomes the amplitude of spin down and vice versa. Then, after the first reflection with the magnitude of spin exchanged, the wave packet returns to the starting position. It turns out that when the wave packet is back superposed centered at the origin, the spin components have a global phase of $\pi$ concerning the initial state, which results in a null fidelity state. However, after two successive reflections, in the new overlapping centered at $j=0$, the spin components have the phase of the initial separable state, which now provides a high-fidelity state. Therefore, we obtain the same results as the CTQW above.

The dynamic evolution from the analytical solution of Eq.~(\ref{Psit}) has a remarkable resemblance with the behavior seen in CTQWs on circular graphs above (see Fig.~\ref{fig.3}, for instance). Suppose we change the DTQW placed on a constrained line to a circular geometry including a periodic boundary such as $j=j\pm L$ instead of $\sigma_x$ gates. In such a case, the wave packet changes its direction, from left to clockwise and from right to counterclockwise. In such a case, the absence of swapping between spin components or phase change makes the superposition always periodically return to the initial state. But now, the initial state also can be found centered at the antipodal position. Therefore, we have the same results of CTQWs shown above transporting a qubit through DTQWs.

\section{Discussion}\label{sec:5}

Quantum walks starting from a delocalized state allow us to recover the initial state with high fidelity for appreciable time scales and after spreading over great distances. The requirement is a smooth and broad enough initial state as a symmetrical Gaussian one. It makes the interference pattern periodically bring the quantum state back to its initial condition when it reaches the opposite and starting points of a closed path, such as the cycle graph. These results lead to the following statements: (i) The predictable behavior on cycle graphs and the gradual decrease of the upper-bound limit of the fidelity over time suggest appropriate time windows to change the graph geometry to achieve dynamic control of such states. (ii) The steady null group velocity of the state differentiates from other proposals to transfer a quantum state. (iii) A quantum walk that promotes a periodic recovery of a quantum state is feasible in the context of quantum circuits. These ideas are the subjects of the following sections. 

\subsection{Quantum walks in dynamic graphs}

\begin{figure}[h!]
\centering
\includegraphics[width=0.6\linewidth]{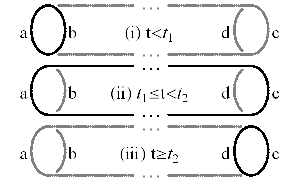}
\caption{Confinement and displacement of a delocalized state: (i) For $t<t_1$: the state initially centered at $a$ is confined inside the small left circular graph (black vertices) between the vertices $a$ and $b$. (ii) At $t=t_1$, when the state has its center at $a$: the left branch of the small left circular graph switches from its right branch to the top and bottom branches of the major graph. It allows the spreading of the state over the outsider major graph (black vertices). (iii) At $t=t_2$, when the state has its center at $c$: the right branch of the small right circular graph switches from the top and bottom branches of the major graph to its left branch. Then, the delocalized state is now inside the small right circular graph (black vertices) confined between $c$ and $d$. Note that the gray vertices indicate where the state cannot spread.}
\label{fig.6}
\end{figure}

Quantum walks can also include the dynamic graphs context, in which a particular sequence of graphs drives the time-evolution of such walks \cite{herrman2019continuous}. Figure~\ref{fig.6} illustrates how we can use the features of quantum walks from delocalized states to confine and transport a quantum state employing dynamic control of circular graphs. First, we time-evolve the delocalized state inside the small left circular graph ($t<t_1$). At a specific time $t_1=2(l+1)\tau_s$ with $l\in\mathbbm{N}$, such that $\tau_s$ corresponds to the transfer time of the small graph, the state has its center at $a$. So, we changed the graph geometry inspired by a railroad switch, redirecting the left branches of the small left circular graph to the outsider graph. It allows the state to follow from $a$ to $c$. When the state has its center at $c$, after a time $t_2=t_1+\tau_e$, where $\tau_e$ is the transfer time from $a$ to $c$, we redirect the right branch of the small right graph to its left branch, confining the state inside it. 

Recently, Vieira et al presented the quantum corralling protocol based on DTQWs. The walk starts from a Gaussian state over an infinite line and evolves as a superposition of two Gaussian wave packets in opposite directions. The authors' protocol leads to the one-directional displacement of the state by changing the quantum gates from Hadamard to $\sigma_x$ at specific time steps and positions \cite{vieira2021quantum}. The main idea of Fig.~\ref{fig.6} could take a qubit as shown here. However, there is a remarkable difference between these two ways to transfer a state. While the corralling protocol allows the transportation of a qubit in wave packets with a finite group velocity, the group velocity remains null all the time here. Furthermore, we should take extra assumptions due to the phase shift between two-level states during the multiple reflections in the corralling \cite{vieira2021quantum,kurzynski2011discrete}, and these assumptions are not necessary for the state transfer with no group velocity once the phase does not change.

\subsection{Quantum walks and spin chains}

The state transfer through spin chains from local states (single spin states) and a superposition of spin states in a wave packet scheme \cite{Bose2007} has been widely addressed in the literature. The displacement of a truncated Gaussian state with high fidelity over a ring of $N$ spins with fixed interactions is possible if the state is sufficiently large, i.e., the superposition takes $L\sim N^{1/3}$ spins. Then, the state travels through the ring with a constant group velocity and negligible dispersion \cite{Osborne2004}. For open-ended spin chains, Ref. \cite{Haselgrove2005} showed how to construct delocalized states at one end of the chain. They evolve to Gaussian states and reach the other end with low dispersion. In such a protocol, Alice and Bob need to access only a single qubit each, but continuous-time control of the interactions of these spins with the lattice is required \cite{Bose2007}.

It is worth noticing that although similar results arise from spin chain protocols, the dynamical evolution shown here is quite different. In the spin chains protocols, the superposition of states has an initial group velocity and travels with an almost soliton-like behavior. So, it is similar to the quantum corralling protocol \cite{vieira2021quantum}. The state transfer here has null group velocity. The state spreads to all vertices of the graph and reemerges on the other end by constructive interference of the wave function. This process does not require control of any interaction on the graph. Such behavior also allows the measure of a high-fidelity qubit when the state superposes centered at initial and antipodal vertices. In these places, the state becomes the initial state product. Therefore, one measurement of the quantum state between these extreme points does not provide faithful information about the qubit. This characteristic could be a resource to transmit securely quantum information, but at the expense of a particular circuital geometry and restricted places to send and receive the quantum state.

\subsection{Quantum walks as quantum circuits}

\begin{figure}[h!]
\centering
\includegraphics[width=0.6\linewidth]{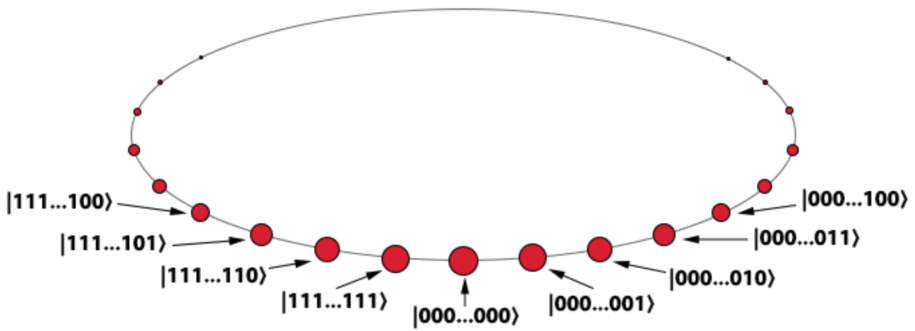}
\caption{For the CTQW, each node of a cycle graph with $n$ vertices is represented by a given basis state, being necessary $\log_2 n$ qubits to represent all vertices.} 
\label{fig.7}
\end{figure}

In this section, we present quantum circuits that are capable of implementing continuous-time and discrete-time quantum walks. Quantum circuits for simulating quantum walks have been proposed \cite{Childs2004thesis,Douglas2009efficient,Slimen2020gen,Portugal2022} as well as their experimental implementation \cite{qiang2016efficient,Acasiete202}. Each vertex of the graph is associated with a basis state, as shown in Fig.~\ref{fig.7} for a circle graph in which the CTQW is implemented. We can employ a similar correspondence between each position of the walker and a basis state. However, the graph geometry must be changed to a straight line. The coin is described by an additional qubit, which one evolves under the action of the Hadamard gate for intermediate steps and NOT gate at the ends of the open-chain. As these circuits are quite demanding in the number of qubits, as well as their depth, their implementation in real quantum devices is limited. For now, only their quantum circuits are being suggested.

The basic ingredient to calculate the fidelity between two quantum states is the SWAP test \cite{Buhrman2001}. Its quantum circuit is presented in Fig.~\ref{fig.8}. At the end of the computation, this circuit returns the probability of measuring the ancilla qubit in the state $\ket{0}$ as $p_0 = \left(1 + \mathcal{F}(t) \right)/2$ or in the state $\ket{1}$ as $p_1 = \left(1 - \mathcal{F}(t) \right)/2$, in which $\mathcal{F}(t)=|\braket{\Psi_{\text{target}}|\Psi(t)}|^2$ is the fidelity between the states $\ket{\Psi(t)}$ and $\ket{\Psi_{\text{target}}}$. This strategy to measure the efficiency of the initial state transfer will be used for both CTQW and DTQW evolutions. The difference between these two kinds of evolutions relies on the quantum circuit to obtain the evolved state $\ket{\Psi(t)}$ and the circuit to obtain the target state must be built case by case.

\begin{figure}[h!]
    \centerline{
            \Qcircuit @C=1em @R=1.5em {
                 \lstick{\ket{0}} & \gate{H}  & \ctrl{1} & \gate{H} & \meter \\
                 \lstick{\ket{\Psi_{\text{target}}}} & \qw & \qswap & \qw & \qw\\
                 \lstick{\ket{\Psi(t)}} & \qw & \qswap \qwx & \qw & \qw
               } }
            \caption{Quantum circuit to calculate the fidelity between the target state $\ket{\Psi_{\text{target}}}$ and the evolved state at time $t$, $\ket{\Psi(t)}$. The first Hadamard gate generates an equal superposed state of the auxiliary qubit, which controls the application of the SWAP gate on the target and evolved states. The second application of the Hadamard gate on the ancilla qubit is necessary to interfere the states, returning the probability $p_0 = \left(1 + |\braket{\Psi_{\text{target}} | \Psi(t)}|^2 \right)/2$ of measuring the first qubit in the state $\ket{0}$ and $p_1 = \left(1 - |\braket{\Psi_{\text{target}} | \Psi(t)}|^2 \right)/2$ the probability of measuring the first qubit in the state $\ket{1}$. }
            \label{fig.8}
\end{figure}

\subsubsection{Quantum circuit for CTQWs}

Now we will introduce the quantum circuits to prepare the evolved and target states. To build the quantum circuit for the CTQW on the cycle graph, first we decompose the evolution operator as in Sect.~\ref{sec:3}, $U(t) = e^{-iAt}=e^{-i\left(F^{\dagger}A_D F\right)t}=F^{\dagger}e^{-iA_D t}F$, where $A_D$ is a diagonal matrix whose elements are given by Eq. (\ref{Eigenvalues_CTQW}) and $F$ is the matrix to implement the quantum Fourier transform. The advantage of this method is that the quantum Fourier transform can be efficiently implemented in polynomial time $O(\log^2 n)$. In contrast, the classical discrete Fourier transform scales exponentially with the number of bits \cite{qiang2016efficient}. The quantum circuit to generate the state $\ket{\Psi(t)}$ from the initial state $\ket{\Psi(0)}$ is described in Fig.~\ref{fig.9}, in which $QFT$ represents the circuit\footnote{The circuit of the quantum Fourier transform can be found in many textbooks of quantum computation. See for instance \cite{nielsen2010quantum}.} of the quantum Fourier transform and $e^{-iA_D t}$ is a diagonal unitary.  
\begin{figure}[h!]
    \centerline{
            \Qcircuit @C=1em @R=.7em {
                  \lstick{\ket{\Psi(0)}} & \gate{QFT} & \gate{e^{-iA_Dt}} & \gate{\left( QFT \right)^{\dag}} & \qw & \rstick{\ket{\Psi(t)}}
               } }
            \caption{Quantum circuit to obtain the evolved state $\ket{\Psi(t)}$ according to the CTQW on the cycle graph starting from an arbitrary initial state $\ket{\Psi(0)}$. $QFT$ is the quantum Fourier transform and the elements of the diagonal matrix $A_D$ are given by Eq. (\ref{Eigenvalues_CTQW}). }
            \label{fig.9}
\end{figure}

The depth of the quantum circuit to evaluate the diagonal unitary operator $e^{-iA_D t}$ by its turn will depend on the dimension of the Hilbert space $n$, as shown in Fig.~\ref{fig.10}. Each phase shift gate is determined by
\begin{equation}
   P(\lambda_j t) = \left( 
        \begin{matrix}
         1 & 0\\
         0 & e^{-i\lambda_jt}
        \end{matrix}  
        \right),
\end{equation}
in which $\lambda_j$ are the eigenvalues of the circulant matrix $A$ given by Eq. (\ref{Eigenvalues_CTQW}). The first phase shift gate $P(\lambda_0 t)$ is activated by the state $\ket{000 \cdots 00}$, the second one $P(\lambda_1 t)$ by $\ket{000 \cdots 01}$ until the last one $P(\lambda_{n-1} t)$ by the state $\ket{111 \cdots 11}$. Although this approach will not be employed here, it is possible to control the precision of each phase $\lambda_j t$ with an additional quantum circuit, since all eigenvalues can be computed efficiently \cite{Childs2004thesis}. Here, the circuit depth $O(n)$ is the most consuming resource since each multi-controlled phase shift gate can be decomposed in $O(\log^2 n)$ one and two qubits gates \cite{nielsen2010quantum}.    
\begin{figure}[h!]
    \centerline{
            \Qcircuit @C=1em @R=1.5em {
               &   & \ctrlo{1}  & \ctrlo{1} & \qw & \cdots & & \ctrl{1} & \ctrl{1} & \qw \\
               &   & \ctrlo{1}  & \ctrlo{1} & \qw & \cdots & & \ctrl{1} & \ctrl{1} & \qw \\
               &   & \ctrlo{1}  & \ctrlo{1} & \qw & \cdots & & \ctrl{1} & \ctrl{1} & \qw \\
               &   &  &  &  &  & &  &  &  \\
               &   & \vdots     & \vdots    &     & \cdots & & \vdots   & \vdots   &     \\
                &   &  &  &  &  & &  &  &  \\
               &   & \ctrlo{1} \qwx{-1}  & \ctrl{1} \qwx{-1}  & \qw & \cdots & & \ctrlo{1} \qwx{-1}& \ctrl{1} \qwx{-1} & \qw \\
        & \lstick{\ket{1}} & \gate{P(\lambda_0 t)}  & \gate{P(\lambda_1 t)} & \qw & \cdots & & \gate{P(\lambda_{n-2} t)} & \gate{P(\lambda_{n-1} t)} & \qw
               } }
            \caption{Quantum circuit to implement the diagonal unitary operator $e^{-iA_D t}$. $P(\lambda_j t) = [1,0;0,e^{-i\lambda_jt}]$ are phase shift gates controlled by the basis states of $\log_2 n$ qubits with $\lambda_j$ being the eigenvalues of the adjacency matrix $A$ (see Eq. (\ref{Eigenvalues_CTQW})) and $t$ is the elapsed time. Open (solid) circle means that the phase gate will be applied if the control qubit is in state $\ket{0}$ ($\ket{1}$).}
            \label{fig.10}
\end{figure}

The initial state of the CTQW can be a localized state, as any basis state vector $\ket{j}$ with $j = \{0,1,\dots,n-1 \}$, or a delocalized state, as a Gaussian state described by Eq. (\ref{Gaussianstate}). Observe that due to the symmetry of the cycle graph, the initial localized state or the center-state of the Gaussian wave packet can be anyone, which will be chosen as the state  $\ket{0}$ ($a=0$). To calculate the fidelity between the target state and the evolved one, we use the SWAP test described in Fig.~\ref{fig.8} considering two cases: (i) the target state $\ket{b}$ is localized, then it will be the following basis state $\ket{\left[n - (n \mod 2)\right]/2}$; (ii) the target state is delocalized, then it is a displaced Gaussian state $\mathcal{D} \ket{\psi_P(0)}$. Therefore, we need to provide quantum circuits to prepare such states. The preparation of a localized target state is quite simple since it is a string of bits, i.e., it can be prepared by the suitable application of $X$ ($NOT$) gates on the state $X^{b_0}X^{b_1}\dots X^{b_{m-1}}\ket{000 \cdots 00}=\ket{b_0b_1b_2 \cdots b_{m-2}b_{m-1}}$, for $b_0, b_1, \cdots b_{m-1} \in \{0,1\}$, which takes at most $m$ gate applications, with $m =\log_2 n$.  

For the preparation of the displaced Gaussian state, first, we build the circuit to implement the displacement operator, which translates the states $\ket{j}$ to $\ket{j+b \mod{n}}$ for $j,b \in \{0,1,\cdots,n-1\}$ through the QFT adders \cite{Perez2017}. The addition modulo $n$ derives from applying the sequence of gates 
\begin{equation}
    IQFT_1.CZ.QFT_1 \ket{j}\ket{b} = \ket{j+b \mod{n}}\ket{b},
    \label{eq:mod}
\end{equation}
where the subindex $1$ indicates the position of the subsystem in which the gate will act on, i.e., the leftmost state. In this situation, the state $\ket{b}$ comprises $m =\log_2 n$ auxiliary qubits initially prepared in one of the basis states $\ket{b_0b_1b_2 \cdots b_{m-2}b_{m-1}}$. As the generalized $CZ$ gate is defined by 
\begin{equation}
CZ\ket{j}\ket{b}= \omega^{jb}\ket{j}\ket{b},
\label{CZ}
\end{equation}
and $QFT_1$ and $IQFT_1$ are given, respectively, by equations (\ref{QFT}) and (\ref{IQFT}), we can demonstrate the validation of Eq. (\ref{eq:mod}), as follows
\begin{eqnarray}
     IQFT_1.CZ.QFT_1 \ket{j}\ket{b} &=& IQFT_1.CZ \frac{1}{\sqrt{n}} \sum_{k=0}^{n-1} \omega^{kj}\ket{k}\ket{b}  \nonumber\\
     &=& IQFT_1 \frac{1}{\sqrt{n}} \sum_{k=0}^{n-1} \omega^{k(b+j)}\ket{k}\ket{b} \nonumber\\
      &=&  \frac{1}{n} \sum_{k,s=0}^{n-1} \omega^{k(b+j-s)}\ket{s}\ket{b} \nonumber\\
      &=&  \sum_{s=0}^{n-1}\left[ \frac{1}{n} \sum_{k=0}^{n-1} \omega^{k(b+j-s)} \right]\ket{s}\ket{b} \nonumber\\
      &=&  \sum_{s=0}^{n-1}\delta_{(b+j \mod{n}), s}\ket{s}\ket{b} \nonumber\\
     &=&
     \ket{j+b \mod{n}} \ket{b}.
\end{eqnarray}
As we already know how to implement the $QFT$ and its inverse, we need to show how to implement the generalized $CZ$ gate, as in Eq. (\ref{CZ}). To implement its quantum circuit, first we write $j = j_0 2^0+j_1 2^1 + \cdots + j_{m-1} 2^{m-1}$ and $b = b_0 2^0+b_1 2^1 + \cdots + b_{m-1} 2^{m-1}$ in binary representation, with $j_s,b_r \in \{0,1\}$ and $s,r = \{0,1,\cdots,m-1\}$. The number $m$ is chosen according to the values of $j$ and $b$. The next step is the multiplication of $j$ and $b$ to compose the phases
\begin{equation}
\omega^{jb} = e^{\frac{i2\pi jb}{n}} = e^{\sum_{r,s=0}^{m-1} i2\pi j_s b_r 2^{s+r-m}}.    
\end{equation}
The contribution of $j$ and $b$ to the phases occurs only for $j_s=b_r=1$ and $s+r<m$, since for $s+r>m$ the phase is an integer multiple of $2\pi$. Therefore, we will apply at most $m(m+1)/2$ controlled rotations $R_{\ell}$, given by  
\begin{equation}
    R_{\ell} = \left(
    \begin{matrix}
        1 & 0 \\
        0 & e^{i2\pi/2^{\ell}} 
    \end{matrix}
    \right),
\end{equation}
where $\ell = \{0,1,...,m-1 \}$. The state $\ket{b}$ controls the action of $R_{\ell}$ over the state $\ket{j}$, as shown in Fig.~\ref{fig.11}.

\begin{figure}[h!]
    \centerline{
            \Qcircuit @C=0.7em @R=1.5em {
               & \lstick{\ket{j_0}}    & \gate{R_{m}}& \gate{R_{m-1}}& \qw & \cdots & & \gate{R_{1}}& \qw           & \qw           &\qw & \cdots & & \qw        & \qw & \cdots & & \qw & \qw       & \qw \\
               & \lstick{\ket{j_1}}    & \qw         & \qw           & \qw & \cdots & & \qw         &\gate{R_{m-1}} & \gate{R_{m-2}}&\qw & \cdots & & \gate{R_1} & \qw & \cdots & & \qw & \qw        & \qw\\
               &                       &             &               &     & \vdots & &             &               &               &    & \vdots & &            &     & \vdots & &     &            &  \\
               & \lstick{\ket{j_{m-1}}}& \qw         & \qw           & \qw & \cdots & & \qw         & \qw           & \qw           &\qw & \cdots & & \qw        & \qw & \cdots & & \qw & \gate{R_1} & \qw \\
               & \lstick{\ket{b_0}}    & \ctrl{-4}   & \qw           & \qw & \cdots & & \qw         & \ctrl{-3}     & \qw           &\qw & \cdots & & \qw        & \qw & \cdots & & \qw & \ctrl{-1}        & \qw\\
                & \lstick{\ket{b_1}}   & \qw         & \ctrl{-5}     & \qw & \cdots & & \qw         & \qw           & \ctrl{-4}     &\qw & \cdots & & \qw        & \qw & \cdots & & \qw & \qw        & \qw\\
               &                       &             &               &     & \vdots & &             &               &               &    & \vdots & &            &     & \vdots & &     &            & \\
               & \lstick{\ket{b_{m-2}}}& \qw         & \qw           & \qw & \cdots & & \qw         & \qw           & \qw           &\qw & \cdots & & \ctrl{-6}  & \qw & \cdots & & \qw & \qw        & \qw \\
               & \lstick{\ket{b_{m-1}}}& \qw         & \qw           & \qw & \cdots & & \ctrl{-8}   & \qw           & \qw           &\qw & \cdots & & \qw        & \qw & \cdots & & \qw & \qw        & \qw      
               } }
            \caption{Quantum circuit to implement the generalized controlled phase gate $CZ$ described in Eq. (\ref{CZ}). The rotations are determined by $R_{\ell} = [1,0;0,\exp{\left(i2\pi/2^{\ell}\right)}]$, where $\ell = \{0,1,...,m-1 \}$ and $m =\log_2 n$, with $n$ being the dimension of the system. 
            }\label{fig.11}
\end{figure}

To prepare an initial state representing a given distribution function, such as the Gaussian state (\ref{Gaussianstate}) or other delocalized states presented in Appendix \ref{sec:apend}, we will use the idea of exact state preparation, as proposed in Ref. \cite{Shende2006} and detailed in the appendix of Ref. \cite{Zanetti2023}. First, let us write a general initial state of $n$ qubits in the binary basis representation 
\begin{equation}
    \ket{\psi_P (0)} = \sum_{j=0}^{n-1} f(j) \ket{j} = \sum_{j_0, j_1, \dots, j_{m-1}=0}^{1} f(j_0, j_1, \dots, j_{m-1}) \ket{j_0, j_1, \dots, j_{m-1}}, 
    \label{eq:estado_inicial_p}
\end{equation}
in which $j = \sum_{k=0}^{m-1} j_k 2^k$. For the distributions we are interested in, $f(j)$ has real positive coefficients satisfying the normalization condition $\sum_{j=0}^{n-1} f(j)^2 =1$. The algorithm for state preparation is the following \cite{Zanetti2023}:
\begin{enumerate}
\item Prepare the $(m-1)$-qubits state
    \begin{equation}
    \ket{\Phi_{m-1}} = \sum_{j_0, j_1, \dots, j_{m-2}=0}^{1} r_{j_0, j_1, \dots, j_{m-2}} \ket{j_0, j_1, \dots, j_{m-2}}, 
    \end{equation}
    with 
    \begin{equation}
        r_{j_0, j_1, \dots, j_{m-2}} = \sqrt{f(j_0, j_1, \dots, j_{m-2}, 0)^2+f(j_0, j_1, \dots, j_{m-2}, 1)^2}.
    \end{equation}
\item Apply $m-1$ multi-controlled unitary gates to the state
    \begin{equation}
        \ket{\Phi_{m-1}}\otimes\ket{0}_{m-1},
    \end{equation}
    where the first $m-1$ qubits control the last qubit through
    \begin{equation}
         \prod_{j_0, j_1, \dots, j_{m-2}=0}^{1} C_{R_{j_0, j_1, \dots, j_{m-2}}}^{0_{j_0}, 1_{j_1}, \dots, (m-2)_{j_{m-2}} \rightarrow (m-1)},
    \end{equation}
    in which 
    \begin{equation}
        C_R^{C_s \rightarrow t}
    \end{equation}
    stands for the application of the unitary gate $R$ on the target qubit $t$. This action is controlled by the qubit $C$ and activated by its state $s$ ($s=\{0,1\}$). Here,
    \begin{equation}
       R_{j_0, j_1, \dots, j_{m-2}} = \exp{\left(-i \frac{\theta_{j_0, j_1, \dots, j_{m-2}}}{2} \sigma_y \right)}
    \end{equation}
    is a rotation around the $y$-axis by an angle 
    \begin{equation}
       \theta_{j_0, j_1, \dots, j_{m-2}} = 2\arctan \left( \frac{f(j_0, j_1, \dots, j_{m-2}, 1)}{f(j_0, j_1, \dots, j_{m-2}, 0)} \right) \label{eq:parameters}
    \end{equation}
    and $\sigma_y = \left( \begin{array}{cc}
       0  & -i \\
       i  &  0
    \end{array} \right) $ is the Pauli matrix. 
\end{enumerate}
We observe that this algorithm must be applied recursively, i.e., the state of $m-1$ qubits $\ket{\Phi_{m-1}}$ will be created by the application of controlled rotations 
\begin{equation}
\prod_{j_0, j_1, \dots, j_{m-3}=0}^{1} C_{R_{j_0, j_1, \dots, j_{m-3}}}^{0_{j_0}, 1_{j_1}, \dots, (m-3)_{j_{m-3}} \rightarrow (m-2)}
\end{equation}
on the state $ \ket{\Phi_{m-2}}\otimes\ket{0}_{m-2}\otimes \ket{0}_{m-1}$, until the rotation on the qubit $0$. This can be better visualized in the quantum circuit presented in Fig.~\ref{fig.12}. 
\begin{figure}[h!]
    \centerline{
            \Qcircuit @C=0.5em @R=1.5em {
               & \lstick{\ket{0}}    & \gate{R_{y}(\xi)}& \ctrlo{1} & \ctrl{1} & \ctrlo{1} & \ctrlo{1} & \ctrl{1} & \ctrl{1}           & \qw           &\qw & \cdots & & \ctrl{1}        & \qw \\
               & \lstick{\ket{0}}    & \qw         & \gate{R_{y}(\theta_0)}          & \gate{R_{y}(\theta_1)} & \ctrlo{1} & \ctrl{1} & \ctrlo{1}         &\ctrl{1} & \qw &\qw & \cdots & & \ctrl{1} & \qw \\
               & \lstick{\ket{0}} & \qw         & \qw           & \qw & \gate{R_{y}(\theta_{00})} & \gate{R_{y}(\theta_{01})} & \gate{R_{y}(\theta_{10})}         & \gate{R_{y}(\theta_{11})}          & \qw           &\qw & \cdots & & \ctrl{1}        & \qw  \\
                &    &   &    &   &  &  &   &   &  &  & \vdots & &  & \\
                & \lstick{\ket{0}}    & \qw   & \qw           & \qw & \qw & \qw & \qw         & \qw     & \qw           &\qw & \cdots & & \gate{R_{y}(\theta_{11 \cdots 1})}        & \qw 
                } }
            \caption{Quantum circuit to prepare a state representing a generalized distribution function with real positive coefficients. The parameters of the rotation around the $y$-axis are determined by Eq. (\ref{eq:parameters}).}
            \label{fig.12}
\end{figure}

We notice that accurate state preparation is resource-intensive since it takes $O(n\log^2 n)$ quantum gates of one and two qubits. Here, we have used the fact that a single-qubit-gate simultaneously controlled by $k$ qubits can be decomposed by $O(k^2)$ one and two-qubit gates \cite{nielsen2010quantum}. Finally, to obtain the circuit to prepare the desired state $\ket{\psi(0)}$, we must analyze case-by-case, i.e., each coefficient must be calculated classically and then replaced in  Eq. (\ref{eq:parameters}) to obtain the $R_y$ gates. Summarizing the quantum complexity of circuit implementation, the overall cost of the protocol to measure the fidelity of quantum state transfer of a CTQW is $O(n\log^2n/\epsilon^2)$, in which $\epsilon$ is the accuracy of measuring the $\sigma_z$ observable of the auxiliary qubit in Fig.~\ref{fig.8}.    

\subsubsection{Quantum circuit for DTQWs}

We explore here the implementation of bounded DTQWs through a quantum circuit. The walker is described by one qubit, initially prepared in the state $\ket{\psi_w (0)}$ (see Eq. (\ref{Psi_0})), while the position states are represented by the basis states of $m$ qubits prepared in the state $\ket{\psi_P (0)}$ (see Eq. (\ref{eq:estado_inicial_p})), as illustrated in Fig.~\ref{fig.13}. We observe that the number of vertices $n$ is efficiently simulated by $m=\log_2 n$ qubits. The preparation of the initial distribution state $\ket{\psi_P (0)} = \sum_j f(j) \ket{j}$ has already been discussed previously, while the preparation of $\ket{\psi_w (0)} = \cos\alpha\ket{0}+e^{i\beta}\sin\alpha\ket{1}$ is made by the application of the gates $P(\beta)R_y(2\alpha)$ on the state $\ket{0}$. 

Along the walk, the Increment gate changes the position of the vertex state $j$ to $j+1$, which one is conditioned to the coin state $\ket{1}$. Initially, the position state $\ket{000...000}$ corresponding to the first vertex $j=0$, changes to $\ket{000...001}$ being the second position ($j=1$), and so on. The Decrement gate, controlled by the coin state $\ket{0}$, changes the initial vertex $j=0$ to the last vertex $j=n-1$, corresponding to the state $\ket{111...111}$, and so on \cite{Douglas2009efficient,Slimen2020gen}. To describe a finite number of steps given by the walker, it is necessary to apply the Increment and Decrement gates for each step. For intermediate steps, the unitary $M$ becomes the Hadamard gate, however, when the walker meets the border vertex, $j= \pm L$, $M=X$. After $n$ and $2n$ recurrently time steps, the superposition states are centered at the antipodal and initial vertices, respectively. The gate complexity of the DTQW is similar to the CTQW case, since the most expensive part of the algorithm is, in general, the quantum state preparation. Each multi-controlled NOT gate used in the Decrement and Increment gates scale as $O(k^2)$, where $k$ is the number of qubits in the gate, if we consider its decomposition in terms of one and two qubits gates. So the cost of Decrement and Increment gates is $O(m^3)$, for $m$ qubits.   
\begin{figure}[h!]
	\centering
	\begin{minipage}{0.45\linewidth}
	        \centerline{
            \Qcircuit @C=0.9em @R=0.5em {
               & \lstick{\ket{0}}  & \gate{\ket{\psi_w (0)}}  &\gate{M} & \qw & \ctrl{1}  & \qw & \ctrlo{1} & \qw \\
               & \lstick{\ket{0}} & \multigate{3}{\ket{\psi_P (0)}} &\qw  & \qw & \multigate{3}{\text{Increment}} & \qw & \multigate{3}{\text{Decrement}} & \qw \\
               & \lstick{\ket{0}} & \ghost{\ket{\psi_P (0)}}   & \qw & \qw   & \ghost{\text{Increment}} & \qw & \ghost{\text{Decrement}} & \qw  \\
               & \vdots & \nghost{\ket{\psi_P (0)}} &\vdots &   & \nghost{\text{Increment}} & \vdots & \nghost{\text{Decrement}} &  \vdots \\
               & \lstick{\ket{0}} & \ghost{\ket{\psi_P (0)}}   & \qw & \qw   & \ghost{\text{Increment}} & \qw & \ghost{\text{Decrement}} & \qw \gategroup{5}{4}{5}{8}{.7em}{_\}}\\
               &&&&&&&&\\
               &&&&&&\text{one discrete time step}&&
             } } 
    \label{subfig.13a}
     \vspace{0.7cm}
	\end{minipage} \\	
	\begin{minipage}{0.45\linewidth}
	    \centerline{
            \Qcircuit @C=0.9em @R=0.5em {
               & \multigate{5}{\text{Increment}} & \qw & & & \qw & \targ & \qw & \qw & \qw & \qw &\cdots & & \qw & \qw \\
               & \ghost{\text{Increment}}        & \qw & && \qw   &  \ctrl{-1}      & \targ & \qw & \qw & \qw  & \cdots & &\qw & \qw   \\
               & \ghost{\text{Increment}}        & \qw  & = &  & \qw   & \ctrl{-1}      & \ctrl{-1} & \targ & \qw & \qw  & \cdots & & \qw & \qw \\
               & \ghost{\text{Increment}}       & \qw & & & \qw   & \ctrl{-1}      & \ctrl{-1} & \ctrl{-1} & \targ & \qw  & \cdots & & \qw & \qw  \\
               & \nghost{\text{Increment}}       &  & & &  &    &       &   &  &  & \vdots   &  &        &   \\
               & \ghost{\text{Increment}}        & \qw & &&\qw   & \ctrl{-1}      & \ctrl{-1} & \ctrl{-1} & \ctrl{-1} & \qw  & \cdots &  & \gate{X} & \qw  
             } }
	\label{subfig.13b}
         \end{minipage}\\
           \vspace{0.7cm}
         \begin{minipage}{0.45\linewidth}
	    \centerline{
            \Qcircuit @C=0.9em @R=0.5em {
               & \multigate{5}{\text{Decrement}} & \qw & & & \qw & \targ & \qw & \qw & \qw & \qw &\cdots & & \qw & \qw \\
               & \ghost{\text{Decrement}}        & \qw & && \qw   &  \ctrlo{-1}      & \targ & \qw & \qw & \qw  & \cdots & &\qw & \qw   \\
               & \ghost{\text{Decrement}}        & \qw  & = &  & \qw   & \ctrlo{-1}      & \ctrlo{-1} & \targ & \qw & \qw  & \cdots & & \qw & \qw \\
               & \ghost{\text{Decrement}}       & \qw & & & \qw   & \ctrlo{-1}      & \ctrlo{-1} & \ctrlo{-1} & \targ & \qw  & \cdots & & \qw & \qw  \\
               & \nghost{\text{Decrement}}       &  & & &  &    &       &   &  &  & \vdots   &  &        &   \\
               & \ghost{\text{Decrement}}        & \qw & &&\qw   & \ctrlo{-1}      & \ctrlo{-1} & \ctrlo{-1} & \ctrlo{-1} & \qw  & \cdots &  & \gate{X} & \qw  
             } }
	\label{subfig.13c}
         \end{minipage}
\caption{(Top) Quantum circuit for the implementation of DTQW \cite{Douglas2009efficient}. Each discrete-time step of the quantum walker is implemented by the coin gate $M$ and by the Increment and Decrement gates referring to the walker position. For intermediate steps the gate $M=H$, but at the borders of the linear graph $j=\pm L$, $M=X$.  The first qubit corresponds to the walker, initially prepared in the state $\ket{\psi_w (0)}$. The computational basis states of the remaining $m$ qubits represent the graph vertices, whose initial state is $\ket{\psi_P (0)}$. (Middle) The Increment gate is controlled by the walker state $\ket{1}$ and corresponds to the displacement between the positions $j \rightarrow j+1$. (Bottom) The Decrement gate is controlled by the walker state $\ket{0}$ and corresponds to the displacement between the positions $j \rightarrow j-1$.}
\label{fig.13}
\end{figure}

\section{Concluding remarks}\label{sec:6}

We have shown that continuous-time and discrete-time quantum walks from a delocalized state on bounded paths could be used as platforms for state transfer. Such walks preserve the fidelity of the state for an appreciable time and over great distances. For this purpose, the state should be smooth and wide enough as the symmetrical Gaussian one, regardless of the graph size. Then, the state displaces without group velocity, and the constructive interference in distinct places leads to periodic recovery of the initial state. We also have outlined an underlying idea based on the dynamic control of these graphs \cite{herrman2019continuous}. It suggests potential applications such as quantum memories \cite{lvovsky2009optical} and sequential transportation of quantum cargo \cite{roy2020how}. Moreover, we have shown one way to implement these walks through quantum circuits. We hope our findings might be helpful to the development of new protocols for quantum communication in the context of quantum walks and spin chains experimental research.

\section*{Acknowledgments}
This work was supported by Conselho Nacional de Desenvolvimento Científico e Tecnológico - CNPq through grant number 409673/2022-6. E. P. M. Amorim thanks J. Longo for her careful reading and corrections of the manuscript. 

\appendix

\section{Other delocalized states} \label{sec:apend}

We choose three more normalized distribution functions to build other delocalized states to compare to the Gaussian state. These distributions are functions of the vertex $j$, have the same initial standard deviation $s$, and each has unique features that slightly differ from the Gaussian distribution.

\begin{figure}[h!]
\centering
\includegraphics[width=0.8\linewidth]{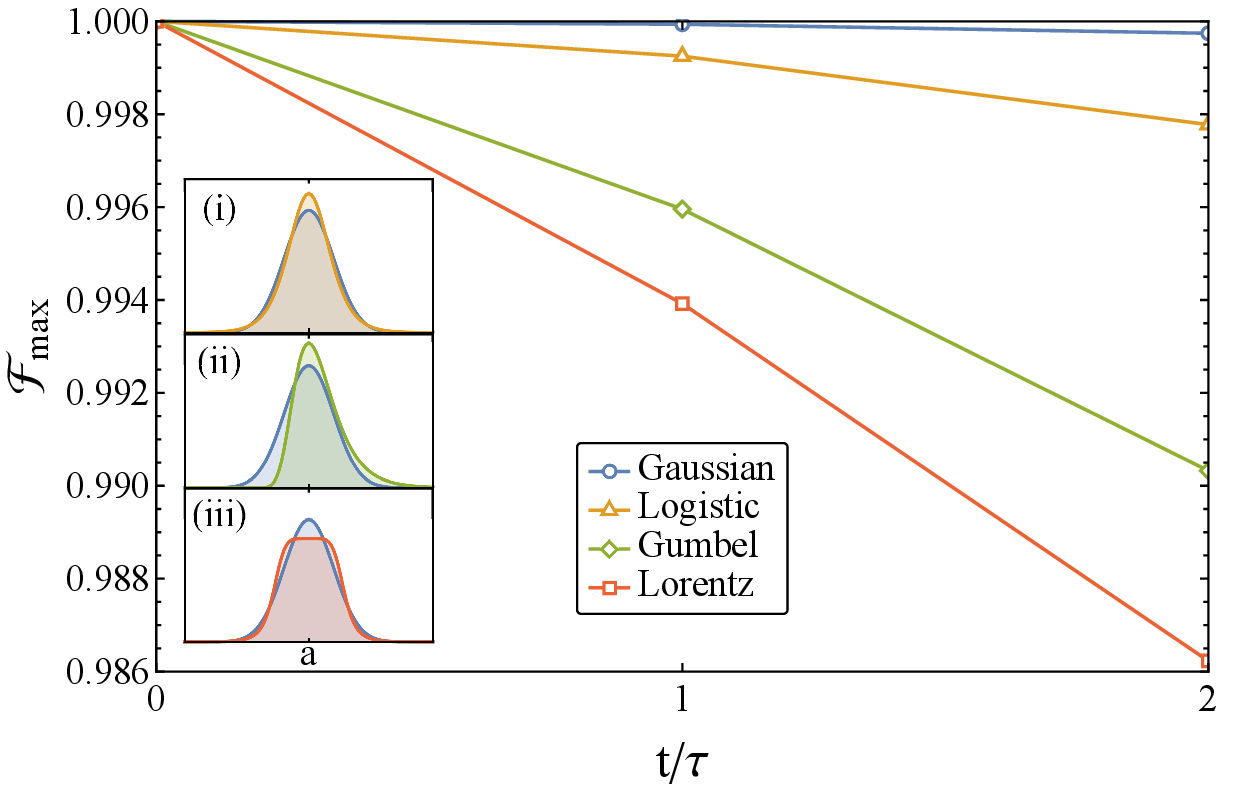}
\caption{Maximum fidelity at $t=\tau$ and $2\tau$ for CTQW over $C_{200}$ starting from delocalized states with $s=10$. Inset: (i) Logistic, (ii) Gumbel, and (iii) Lorentz-like distributions together with the Gaussian one at $t=0$ for comparison. The lines between points just guide the eyes.}
\label{fig.14}
\end{figure}

The first distribution is a logistic one,
\begin{equation}
|f_{\text{Log}}(j)|^2 = \frac{\alpha}{s}\text{sech}^2\left(\frac{2\alpha}{s}j\right),
\label{f_Log}
\end{equation}
with $\alpha=\pi/(4\sqrt{3})$. This function has a Gaussian-like shape with a higher kurtosis. The second one is the Gumbel distribution, 
\begin{equation}
|f_{\text{Gum}}(j)|^2 = \frac{\beta}{s}\exp\left(-\frac{\beta}{s}j-e^{-\frac{\beta}{s}j}\right),
\label{f_Gum}
\end{equation}
such that $\beta=\pi/\sqrt{6}$. It has a positive skew, different from the symmetrical Gaussian distribution. The last one is a Lorentz-like distribution, 
\begin{equation}
|f_{\text{Lor}}(j)|^2=\gamma\frac{s^5}{j^6+8s^6},
\label{f_Lor}
\end{equation}
with $\gamma=6\sqrt{2}/\pi$. This distribution has a rounded plateau, is smoother than the others, and the atypical $j$-exponent yields this function to be normalized. Figure~\ref{fig.14} shows the maxima of the fidelity corresponding to the first state transfer $\tau$ and a complete period $2\tau$ to check the periodicity. The inset shows a comparison between each distribution with the Gaussian one. Since both heavy-tailed and positively skewed states exhibit higher fidelity than the low smoothness state, we also consider a truncated uniform state to examine the role of smoothness,
\begin{equation}
\ket{\Psi(0)}=\sum_{k=-a}^{a} \frac{1}{\sqrt{2a+1}} \ket{k+a},
\label{Psi0_Uniform}
\end{equation}
where $s^2=(a^2+a)/3$. For this case, the fidelity dropped to about $0.90$ and $0.88$, respectively, at $t=\tau$ and $2\tau$. Moreover, since $a$ should be an integer, $s$ is overestimated in $1\%$.

\end{document}